\documentclass[3p, onecolumn]{elsarticle}
\usepackage{hyperref,color}
\usepackage[displaymath, mathlines]{lineno}
\modulolinenumbers[1]
\usepackage{todonotes}
\usepackage{ulem}
\usepackage{amssymb}
\usepackage{booktabs}
\usepackage{chemformula}
\usepackage{gensymb}
\usepackage{comment}
\usepackage{svg}
\setchemformula{radical-radius=1pt}
\RequirePackage{xspace}
\DeclareFontFamily{OT1}{pzc}{}
\DeclareFontShape{OT1}{pzc}{m}{it}{<-> s * [1.200] pzcmi7t}{}
\DeclareMathAlphabet{\mathpzc}{OT1}{pzc}{m}{it}
\usepackage{amsfonts}
\usepackage{textcomp}
\journal{Nuclear Instruments and Methods A}
\usepackage{graphicx}
\usepackage{caption}
\usepackage{subcaption}

\usepackage{multirow}
\newcommand{\sipm}{SiPM}
\newcommand{\sipms}{SiPMs}
\newcommand{\cherenkov}{Cherenkov}
\newcommand{\DR}{DR}
\newcommand{\bgo}{BGO}
\newcommand{\pbwo}{\ensuremath{\rm PbWO_4}}
\newcommand{\pbf}{\ensuremath{\rm PbF_2}}

\begin{document}

\begin{frontmatter}

\title{Studies of Cherenkov Photon Production in PbF$_2$ Crystals using Proton Beams at Fermilab}

\author[uva]{Thomas Anderson}
\author[umd]{Alberto Belloni}
\author[fnal]{Grace Cummings}
\author[umd]{Sarah Eno}
\author[howard]{Nora Fischer}
\author[umich]{Liang Guan}
\author[umich]{Yuxiang Guo}
\author[uva]{Robert Hirosky}
\author[fnal]{James Hirschauer}
\author[umd]{Yihui Lai}
\author[umich]{Daniel Levin}
\author[umich]{Hui-Chi Lin}
\author[umd]{Mekhala Paranjpe}
\author[umich]{Jianming Qian}
\author[umich]{Bing Zhou}
\author[umich]{Junjie Zhu\corref{cor}}
\ead{junjie@umich.edu}
\author[caltech]{Ren-Yuan Zhu}

\cortext[cor]{Corresponding author}

\affiliation[uva]{ organization={Department of Physics, University of Virginia},city={Charlottesville},state={VA},country={USA}}
\affiliation[umd]{ organization={Department of Physics, University of Maryland},city={College Park},state={MD},country={USA}}
\affiliation[fnal]{ organization={Fermi National Accelerator Laboratory}, city={Batavia}, state={IL}}
\affiliation[howard]{ organization={Department of Computer Engineering, Howard University},city={Washington},state={DC},country={USA}}
\affiliation[umich]{  organization={Department of Physics, University of Michigan}, city={Ann Arbor}, state={MI}, country={USA}}
\affiliation[caltech]{ organization={Department of Physics, Caltech},city={Pasadena},state={CA},country={USA}}

\begin{abstract}
Future lepton colliders such as the FCC-ee, CEPC, ILC, or a muon collider will collect large data samples that allow precision physics studies with unprecedented accuracy, especially when the data is collected by innovative state-of-the-art detectors.  An electromagnetic calorimeter based on scintillating crystals, designed to separately record \cherenkov\ and scintillation light, can achieve precision measurements of electrons and photons without sacrificing jet energy resolution, given adequate light collection efficiency and separation. This paper presents initial measurements from a program aimed at developing such a calorimeter system for future colliders. We focus on using PbF$_2$ crystals to enhance the understanding of Cherenkov light collection, marking the first step in this endeavor.
\end{abstract}


\end{frontmatter}

%\linenumbers
\section{Introduction}
\label{sec:intro}
The physics programs of future lepton colliders can be enhanced through a calorimeter system that offers state-of-the-art energy resolutions for both hadrons and photons/electrons (``EM" objects)~\cite{FCC:2018byv, CEPCStudyGroup:2018ghi, ILC:2013jhg, Black:2022cth}. Calorimetric measurements of hadron energies have a resolution component arising from large event-by-event fluctuations in the fraction of ``invisible energy" due to nuclear binding energies, which often do not produce detectable signals. There are three main thrusts towards improving hadronic resolution: high granularity calorimetry with particle flow (which aims to utilize the calorimeter only for neutral hadrons, while relying on tracking detectors for charged particle momentum measurement)~\cite{Sefkow:2015hna}, compensating hadronic calorimetry (which enhances processes such as neutron interactions correlated with nuclear breakup but which are only effective for relatively low sampling fractions)~\cite{Acosta:1991ap}, and dual readout (\DR) calorimetry (which uses proxies to estimate the ``invisible energy'' in hadron showers)~\cite{RevModPhys.90.025002}. The first two methods, along with the most common implementation of \DR\ that uses plastic and quartz (or clear) fibers embedded in an absorber to produce and collect signals from scintillation and Cherenkov radiation separately, are based on sampling calorimeters. These approaches suffer from a reduced resolution for EM objects, limited to approximately $13\% / \sqrt{E}$~\cite{RevModPhys.90.025002}.  

The reduced EM energy resolution can be overcome by incorporating a homogeneous electromagnetic calorimeter made of scintillating crystals with \DR\ capabilities~\cite{Lucchini:2020bac}.  These crystals produce both scintillation and \cherenkov\ light. Cherenkov light is produced only by particles traveling faster than the speed of light in the crystal, whereas scintillation light is produced by all charged particles. By measuring each type of light separately, it is possible to estimate the fraction of the EM component and the total energy on an event-by-event basis. The \DR\ method can overcome the typically poor hadronic resolutions of calorimeters with
crystal EM calorimeters, such as seen in the CMS detector~\cite{CMS:2008xjf}.

Pioneering work on \DR\ crystal EM calorimetry was done by the DREAM/RD52 collaboration~\cite{Cascella:2013jka, RevModPhys.90.025002}. However, at that time, it was not possible to get sufficient \cherenkov\ photons without sacrificing the precision EM resolution provided by the high scintillation light output of the crystals used. 
In particular, their work assumed one photodetector per crystal and used different optical filters and the arrival time of the photons to distinguish \cherenkov\ photons from scintillation photons. To obtain a clear early peak from the \cherenkov\ photons, they had to suppress the scintillation light to a degree that photostatistics dominated its measurement, leading to EM energy resolutions comparable to those of sampling calorimeters.

Recent innovations in photodetector technology could overcome these difficulties. Modern silicon photomultipliers (\sipms) offer extended wavelength sensitivity below 400 nm and above 600 nm, enabling high collection efficiency for \cherenkov\ photons. Their relatively low price allows for the use of multiple \sipms\ to read out each crystal, increasing the photosensitive area coverage.  
In this article, we present measurements of Cherenkov photons using \sipms\ for \pbf, a non-scintillating crystal with an index of refraction similar to candidates for such a calorimeter, including \pbwo\ and \bgo. 

\section{Experimental setup}
\label{sec:testbeam}
\begin{figure}[bt]
  \centering
    \subfloat[]{\includegraphics[width=0.45\textwidth]{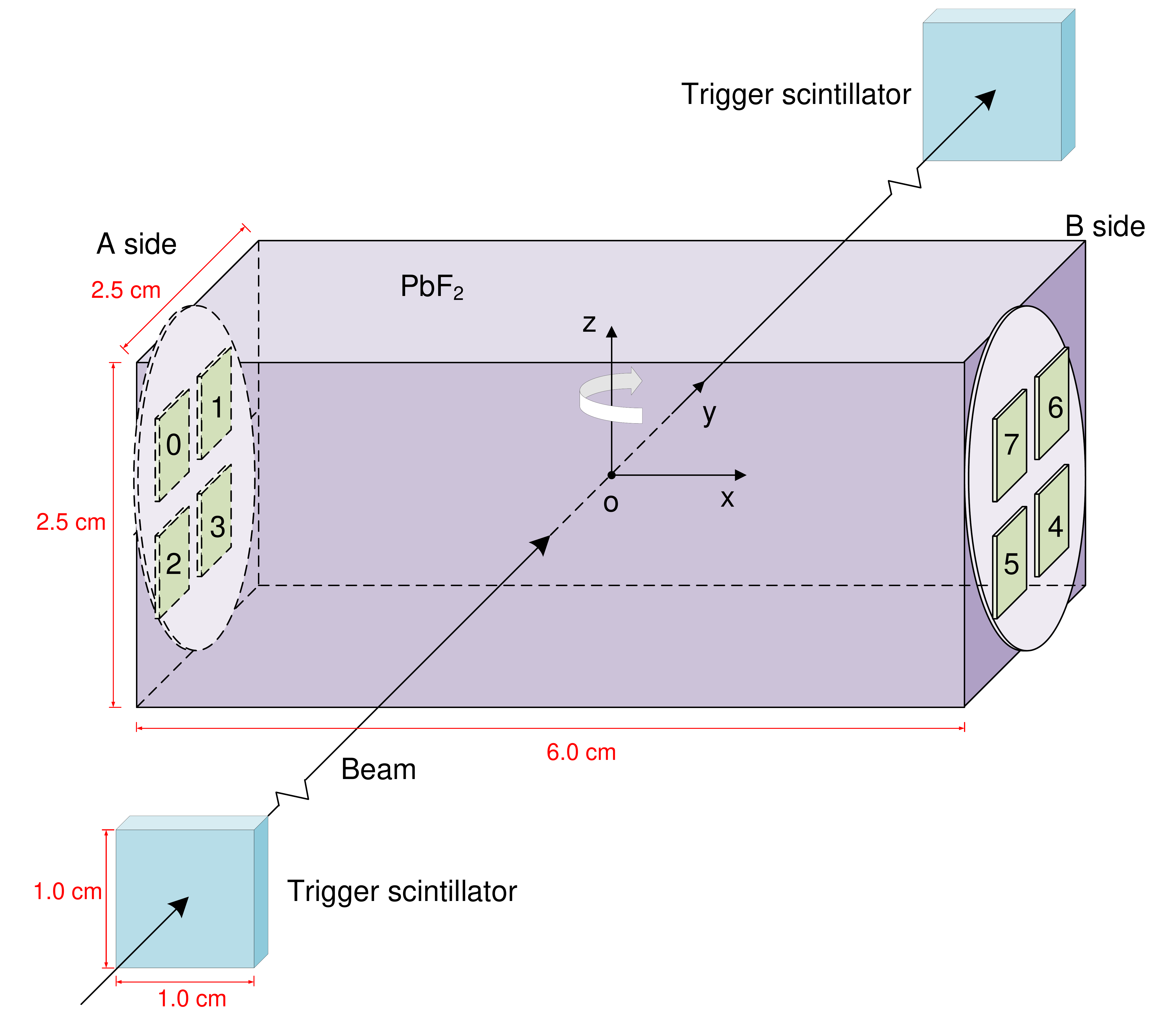} \label{fig:crystal-dimension}} \\
    \subfloat[]{\includegraphics[width=0.45\textwidth]{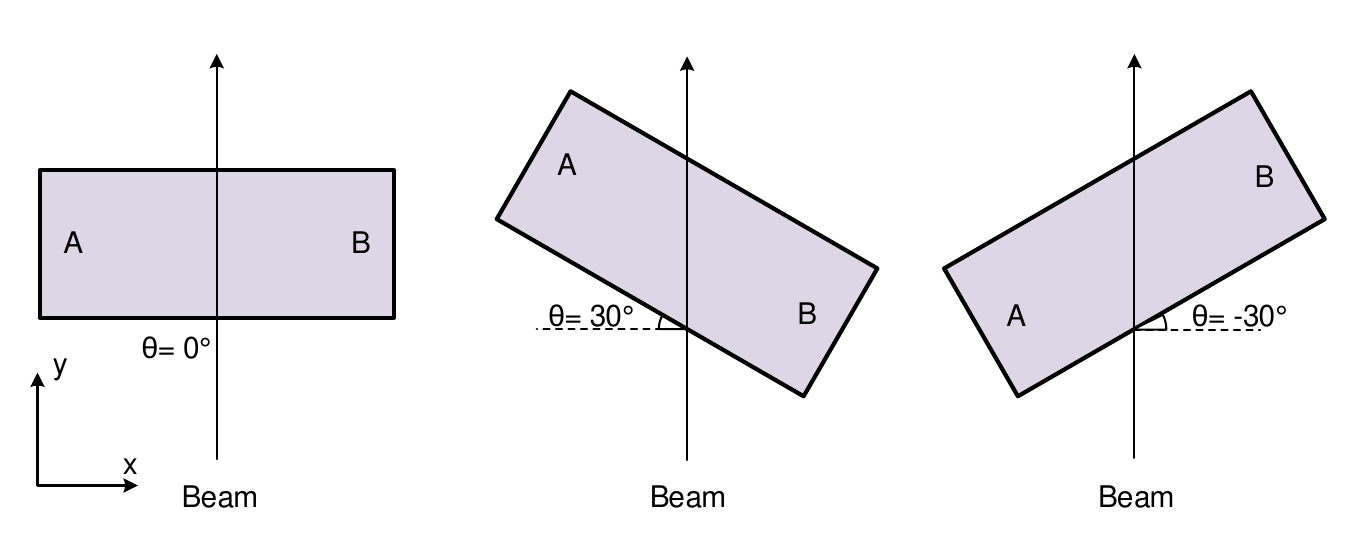} \label{fig:tb-rotation} }
    \caption{(a): Experimental setup; and (b) definition of the rotation angle. See text for details.}
  \label{fig:tb_setup_Fermilab}
\end{figure}

The data was collected in June 2023 using 120 GeV proton beams at Fermilab's test beam facility~\cite{ftbf}. The beam produced one spill per minute, with approximately 45k protons in the first four seconds. The beam had a spread of 8 mm by 4 mm. Figure~\ref{fig:crystal-dimension} illustrates the schematic layout of the experimental setup. Coincidence signals from two small scintillation tiles ($10 \times 10$ mm$^2$), placed 34 cm apart, were used to trigger the data acquisition system. Detailed descriptions of the setup are listed below. 

\textbf{Crystal}: The \pbf\ crystal used in the experiment was purchased from the Shanghai Institute of Ceramics, Chinese Academy of Sciences. It has dimensions of $25 \times 25 \times 60$ mm$^3$, a density of 7.77 g/cm$^3$, and an index of refraction of $1.82$ at 450\,${\rm nm}$~\cite{Allison:2016lfl}. The corresponding Cherenkov angle is $\theta_c=56.7^\circ$ and the total internal reflection angle from the crystal to air is $33.3^{\circ}$. Additionally, the radiation length is 0.93 cm and the nuclear interaction length is 21.0 cm for PbF$_2$~\cite{ParticleDataGroup:2022pth}.  

\textbf{\sipms}: \sipms\ (S14160-6050HS~\cite{sipm_data_sheet}) from the Hamamatsu Corporation were used to read out the raw signals from the crystal. Each \sipm\ module features an effective photosensitive area of $6 \times 6$ mm$^2$, with a micro-cell pitch of 50 $\mu$m and a total of 14,331 cells per module, resulting in a geometrical fill factor of 74\%. These modules are of surface mount type and include a silicone window with a refractive index of 1.57. At 25$^\circ$C, the \sipms\ have a photon detection efficiency of 50\% at 450 nm~\cite{sipm_data_sheet}. However, this efficiency decreases to around 25\% at 300 nm and further decreases for wavelengths below 300 nm. Similarly, the efficiency decreases from 50\% to 10\% as the wavelength increases from 600 nm to 800 nm. The \sipms\ were powered with an overvoltage of 2.7 V. 

Each end of the crystal was equipped with four \sipms, as depicted in Fig.~\ref{fig:tb_setup_Fermilab}. This setup provides an effective photosensitive area of $4 \times 6 \times 6=144$ mm$^2$ at each end, accounting for roughly 23\% of the crystal's cross-sectional area. The four \sipms\ on side A were designated as channels 0, 1, 2 and 3, while those on side B were labeled as channels 4, 5, 6, and 7. The positions of each channel are also illustrated in Fig.~\ref{fig:crystal-dimension}. 

\textbf{Silicone rubber}: The \sipms\ were coupled to the two crystal surfaces using a silicone rubber manufactured by the Eljen Technology. This rubber has a refractive index of 1.43 and a thickness of 1.5 mm~\cite{silicon_cookie}. The total internal reflection angle from the crystal to the rubber is $51.8^{\circ}$.  

\textbf{Wrapping material}: The four crystal surfaces not coupled with the \sipms\ were instead wrapped with Teflon tape, known for its diffuse light reflection properties. These crystal surfaces were thoroughly cleaned before applying three layers of Teflon tape to each surface.

\textbf{Rotating platform}: The crystal was mounted on an HT03RA100 precision motorized rotating platform, allowing rotation along the $z$-axis within a range of $\pm 90^{\circ}$. The rotation angle ($\theta$) is defined as $0^{\circ}$ when the beamline is perpendicular to the crystal's long surface. Positive angles indicate clockwise rotation along the $z$-axis, while negative angles indicate counter-clockwise rotation, as illustrated in Fig.~\ref{fig:tb-rotation}.

The path length of a proton inside the crystal ($\ell$) varies with $\theta$. For $\theta<67.4^{\circ}$, $\ell$ increases with $\theta$ according to $\ell=2.5/\cos\theta$ cm. However, for $\theta>67.4^{\circ}$, $\ell$ decreases with increasing $\theta$ following $\ell=6/\sin \theta$ cm. The maximum path length of 6.5 cm occurs at $\theta=67.4^{\circ}$, corresponding to the proton beam entering through the crystal's diagonal. 

\begin{figure}[bt]
    \centering
    \includegraphics[width=0.38\textwidth]{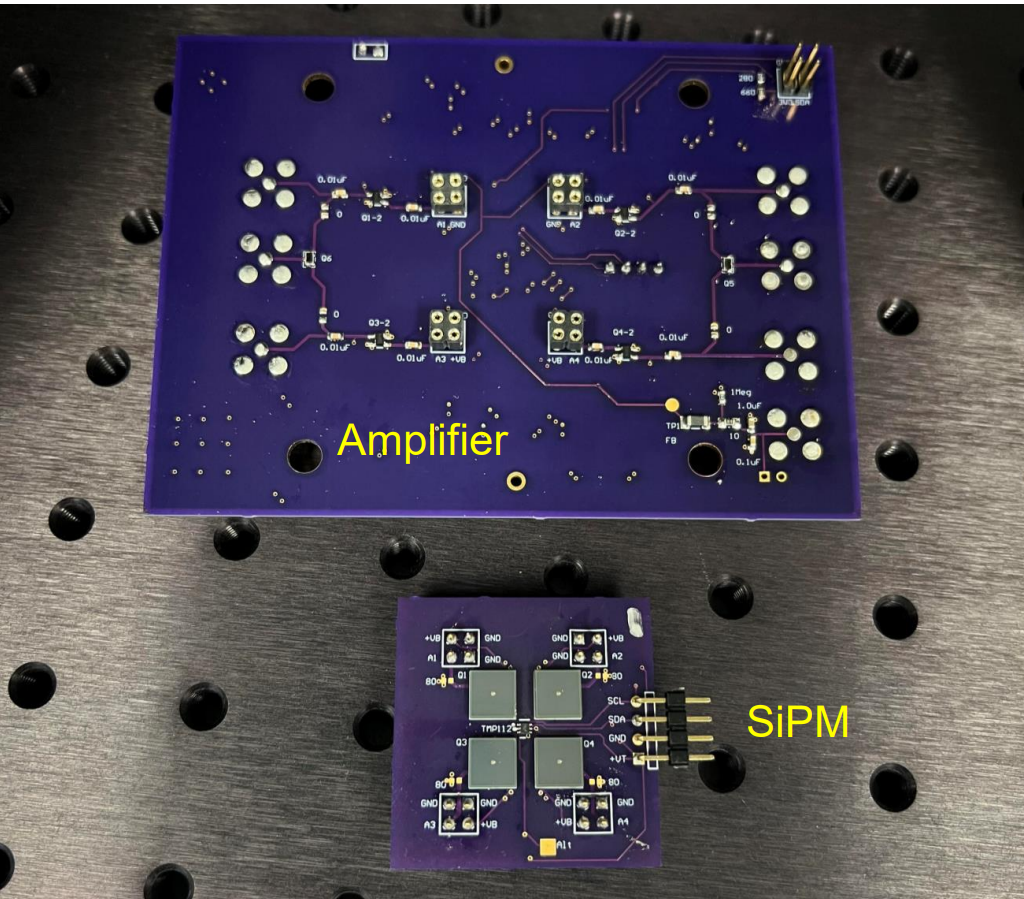}
    \caption{Top: the board that amplifies the \sipms\ output signal; Bottom: the adapter board with four \sipms\ mounted.}
    \label{fig:vaampboard}
\end{figure}

\begin{figure}[bt]
    \centering
    \includegraphics[width=0.50\textwidth]{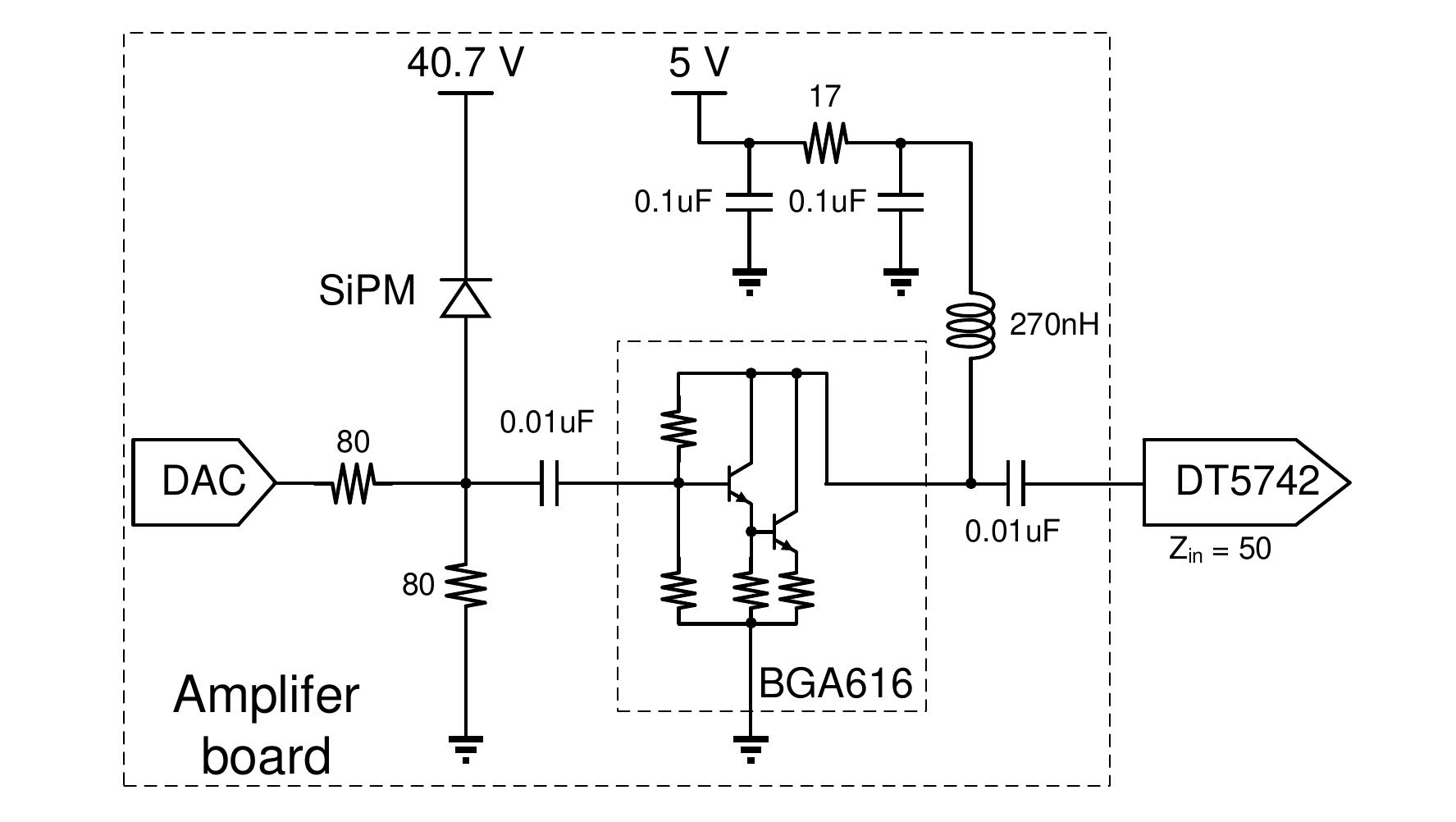}
    \caption{Schematic of one readout channel.}
    \label{fig:vaschematic}
\end{figure}

\textbf{\sipm\ amplifier board}: The raw signals from the \sipms\ were processed through an amplifier front-end board, as depicted in Fig.~\ref{fig:vaampboard}. A small adapter board with four \sipms\ mounted is also shown at the bottom of the same figure. The amplifier board not only supplies voltages to the \sipm\ adapter board but also amplifies the raw signal. A schematic of this amplifier board is presented in Fig.~\ref{fig:vaschematic}, where the bias voltage for each \sipm\ is independently adjustable. Amplification is achieved using the Infineon RF amplifier BGA616, which features a 3 dB-bandwidth from 0 to 2.7 GHz and provides a typical gain of 19.0 dB at 1.0 GHz. The amplified signal is AC-coupled to the digitizer.

The SiPM response was calibrated on the bench using a highly attenuated laser pulse to measure the distribution of the first
several photoelectron (PE) peaks and to determine ADC-to-PE conversion.  An additional gain stage ($\sim 10 \times$) was included to ensure a 1 PE signal $>25$ ADC  counts for this measurement.  After correcting for the low gain configuration used in these studies the response of
0.7 mV/PE corresponded to approximately 3 ADC counts/PE.

\textbf{Trigger and data acquisition system}: Two scintillation tiles, each with an area of $ 10 \times 10$ mm$^2$, were positioned along the path of the proton beam to provide a coincidence trigger signal for the data acquisition system. These two small scintillators were placed 34 cm apart so that the events recorded have the incident proton's angle within $\pm 1.7^{\circ}$. The signals from these tiles were initially detected by two \sipms, amplified and then processed through a discriminator module and a coincidence module. 

\begin{figure}[bt]
  \centering
    \includegraphics[width=0.8\textwidth]{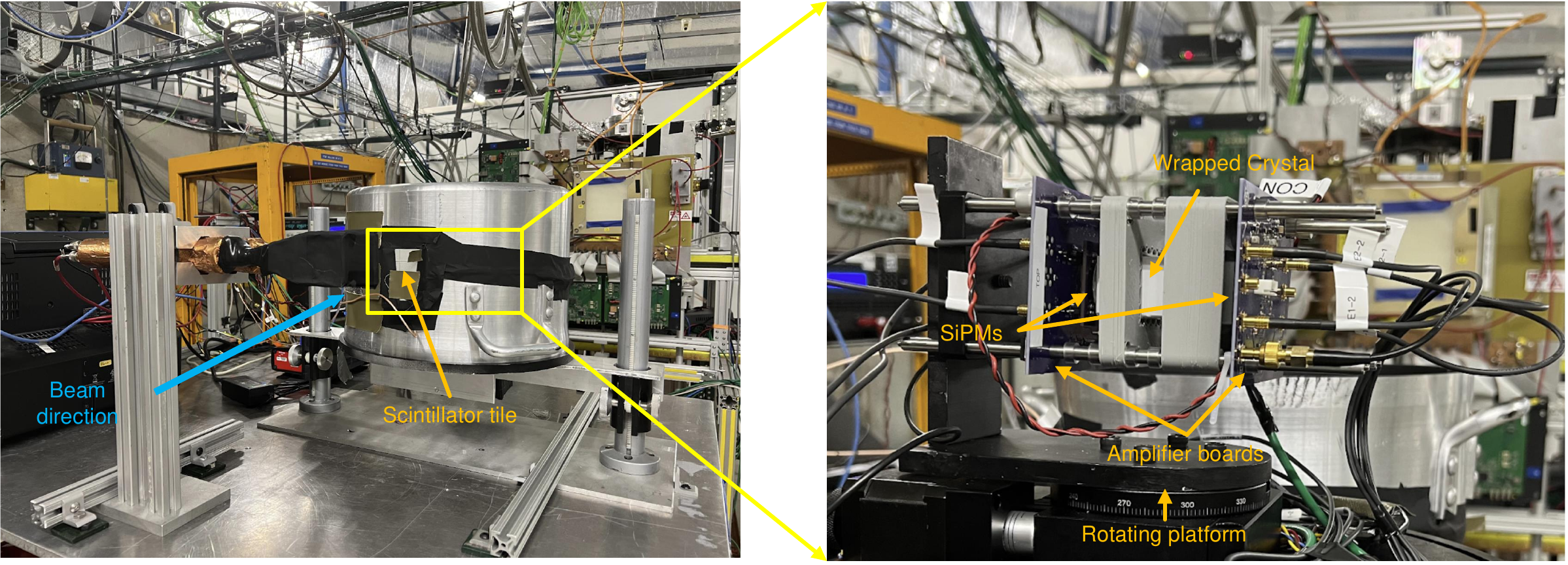} \label{fig:tb-setup1}
    \caption{Actual experimental setup at Fermilab. The entire assembly was enclosed within a steel pot with two scintillation tiles mounted on the pot's surface.}
  \label{fig:testbeam_setup}
\end{figure}

\textbf{Actual setup}: A photograph depicting the actual experimental setup along the beamline is shown in Fig.~\ref{fig:testbeam_setup}. To secure the crystal and boards, a 60 mm optical cage system from Thorlabs, comprising four rigid steel rods and two cage plates, was used. In addition, two plastic holders were positioned in the center to prevent the crystal from dislodging. This entire assembly was enclosed within a steel pot, serving as both a black box and a Faraday cage to shield against electromagnetic pulses. A thick two-layer curtain was employed to envelop the entire setup, preventing outside light from entering. The two scintillation tiles were mounted on the pot's surface to provide the trigger signal, and both the scintillation tiles and the crystal were aligned with the beamline.

The angle between the beam direction and the normal to the crystal's long surface was systematically altered. The angular definition is provided in Fig.~\ref{fig:tb-rotation}. Data were collected for angles ranging from $-90^{\circ}$ to $+90^{\circ}$ in increments of $10^{\circ}$.

\section{Simulation description}
\label{sec:simulation}
To gain a better understanding of the acquired data, we conducted detailed detector Monte Carlo (MC) simulation studies using {\textsc{geant4}}~\cite{GEANT4:2002zbu}. The simulation framework utilizes the {\textsc{geant4}} wrapper dd4hep~\cite{dd4hep}, which can be accessed on GitHub~\cite{mekhalacode}. The settings for the optical properties of the wrappers and interfaces significantly impact the degree of agreement between our data and the simulation results.

Materials included in the simulation model are PbF$_2$ for the crystal, silicone for the rubber, \sipms, and the silicone resin window enveloping each \sipms. Key properties considered include the refractive indices of the crystal, the silicone rubber coupler, and the silicone resin window, as well as the transmission spectrum of the crystal across different wavelengths. We treat the \sipms\ as optical mediums rather than electronic ones, setting their refractive index to match to that of the adjacent volume (i.e. the resin) to minimize total internal reflections. The wavelength-dependent quantum efficiency is incorporated into the post-simulation processing script. 

Photon counting in the script is done both at the generation level and at the \sipms. Each photon track is characterized by a sequence of steps, defined by the G4step class, which includes starting and ending coordinates. The track object provides information about the material of the current step. Photons generated in each medium are recorded at their initial step. For \sipms, photons are terminated inside \sipms\ and counted for each channel.  This counting is based on whether the process is Cherenkov emission or not. Additional conditions are applied, such as discarding photons produced inside the \sipms\ and only considering tracks that hit the photosensitive surface of the \sipms.

For the comparison with experimental data, 10k proton events are simulated for each rotation angle. All events are used to calculate the most probable value of photoelectron counts. Since the wrapped crystal does not have perfectly reflecting faces, the long faces are assigned a {\textsc{geant4}} BorderSurface of ground dielectric-metal. This surface is considered rough, and the proportions of specular spike, specular lobe, backscatter, and Lambertian reflections need to be adjusted. We adjust the percentages of specular lobe and specular spike types, while keeping the backscatter and Lambertian types fixed at 0\%. With this notation, a perfectly smooth surface corresponds to a 100\% specular spike parameter setting. In our default simulation, we set the lobe and spike reflection types to be 25\% and 75\%, respectively. 

\section{Results}
\label{sec:results}
The average trigger rate for the first four seconds of each minute was about 1 kHz, and data were collected for about 15 minutes at each rotation angle. Approximately 40k to 70k events were recorded for each rotation angle. Events with any saturated readout were discarded. The fraction of discarded events depends on the rotation angle: $\sim 2$\% for $|\theta|<30^{\circ}$, $\sim 10$\% for $30^{\circ}<|\theta|<60^{\circ}$, and $\sim 30$\% for $|\theta|>60^{\circ}$. Larger rotation angles result in longer path lengths and enhance Cherenkov photon production. 

The raw signals for a typical event are displayed in Fig.~\ref{fig:raw_spectrum} for two \sipm\ channels (0 and 7). The pedestal, defined as the average ADC value over the first 15 ns, is subtracted from the entire timing spectrum. Fluctuations in the first 15 ns indicate the average noise level across the system. To eliminate timing variations of the DRS trigger channel, event-by-event timing alignment is performed by aligning the leading edge of the trigger channel and applying the same time offset to all readout channels of that event. The resulting spectrum, shown in Fig.~\ref{fig:processed_spectrum}, is then fitted using the following function:
\begin{figure}[bt]
    \centering
         \subfloat[]{\includegraphics[width=0.40\textwidth]{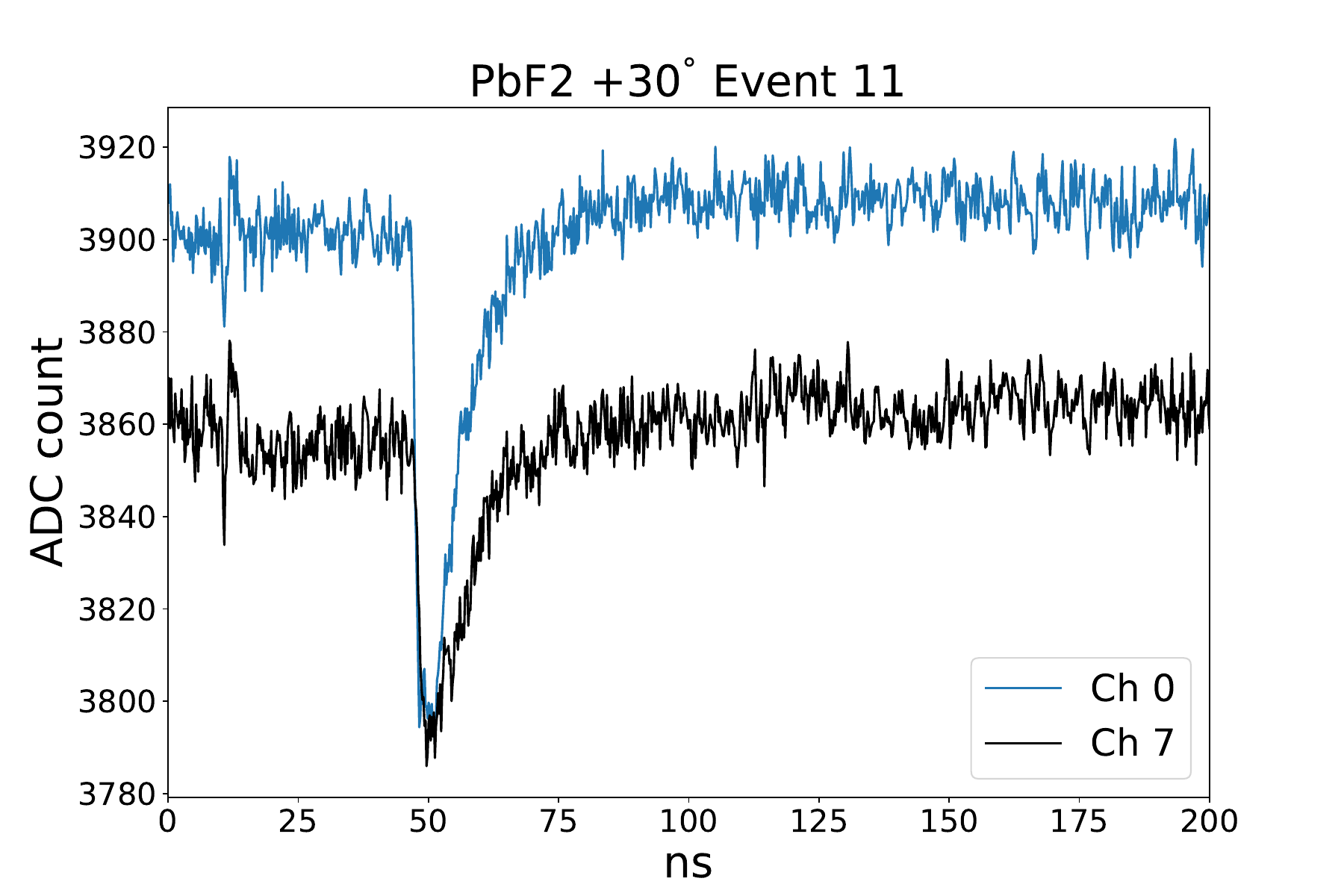} \label{fig:raw_spectrum} } 
         \subfloat[]{\includegraphics[width=0.40\textwidth]{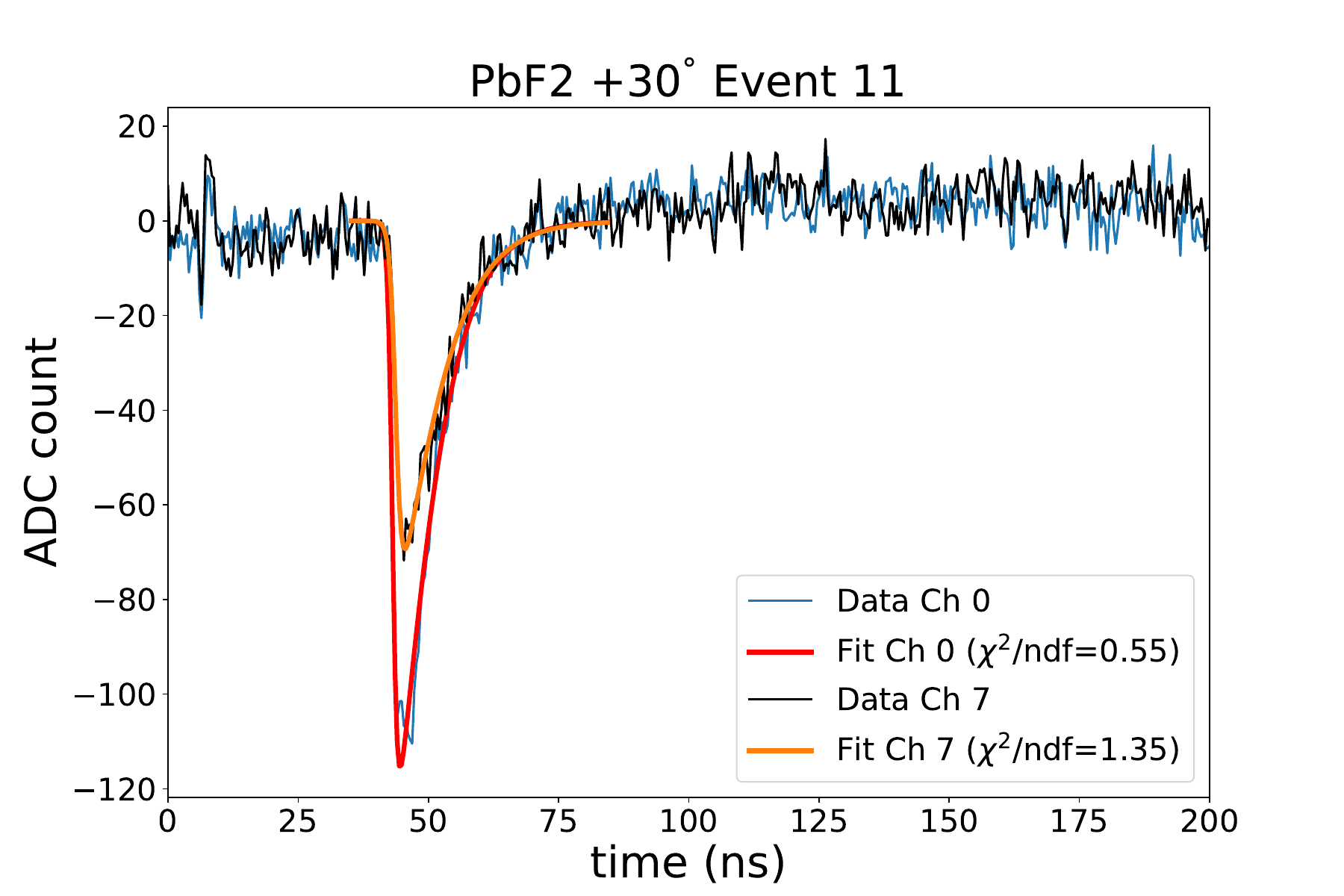} \label{fig:processed_spectrum} } 
    \caption{(a) Raw timing spectrum for one event with two channels; and (b) processed timing spectra with fitted curves overlaid.}
    \label{fig:event_timing_spectrum}
\end{figure}

\begin{equation}
y(t) = A \times \frac{1}{e^{-\frac{t-t_0}{\tau_r}}+1} \times \frac{1}{e^{\frac{t-t_0}{\tau_d}}+1}.
\label{eqn:fit_eqn}
\end{equation}

The fitting function used has four parameters: $A$, $t_0$, $\tau_r$, and $\tau_d$. Here, $t=t_0$ corresponds to the time at the peak position, $\tau_r$ and $\tau_d$ represent the rising and decaying time constants, and $A/4$ is the amplitude at the peak position. 

Events must have a peak amplitude at least five times higher than the noise level. The timing spectrum for each channel is then fitted using the function described in Eqn.~\ref{eqn:fit_eqn}. To ensure a reasonable fit, the fitted peak amplitude must agree with the pedestal-subtracted raw peak amplitude within $\pm 10$ ADC counts. In addition, the fitted $\chi^2$ must be less than $1.5$ (for signals with a peak amplitude above 70 ADC counts) or 9 (for signals with a peak amplitude below 70 ADC counts). The fitted curves for the raw signals of these two channels are shown in Fig.~\ref{fig:processed_spectrum}.  

The response of the eight \sipms\ is examined by analyzing the average timing spectrum for all events collected at $\theta=0^{\circ}$. Channel 2 exhibits a lower response compared to the other seven channels. This discrepancy is attributed to possible air gaps introduced between that \sipm\ and the silicone rubber during the assembly process. Subsequent investigations did not reveal any issues with that \sipm, so its outputs are excluded from the study. The average timing spectra for the remaining seven channels are consistent with each other, indicating uniform responses for these \sipms.

\begin{figure}[bt]
    \centering
    \subfloat[]{\includegraphics[width=0.35\textwidth]{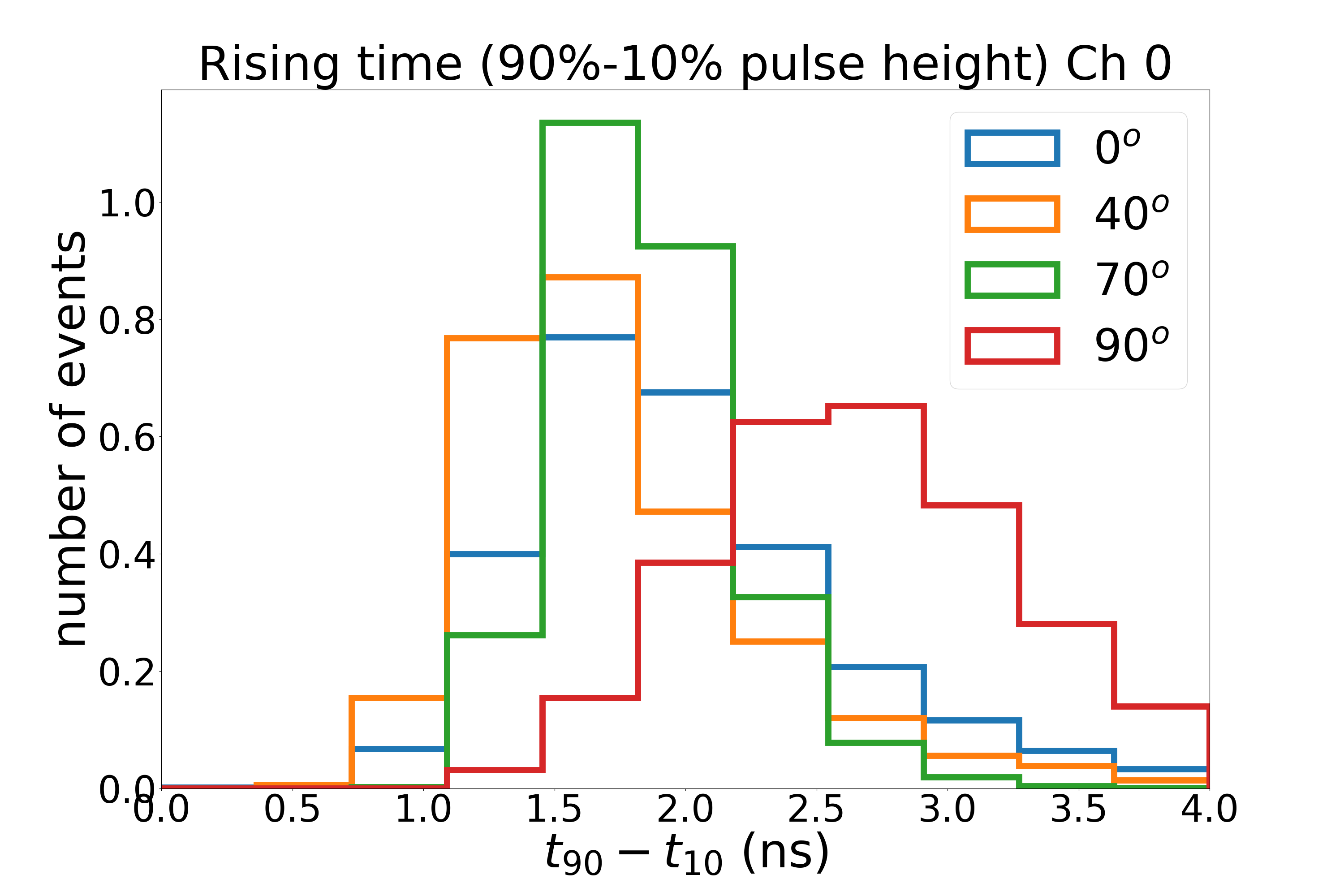} \label{fig:rising_time_ch0} } 
    \subfloat[]{\includegraphics[width=0.35\textwidth]{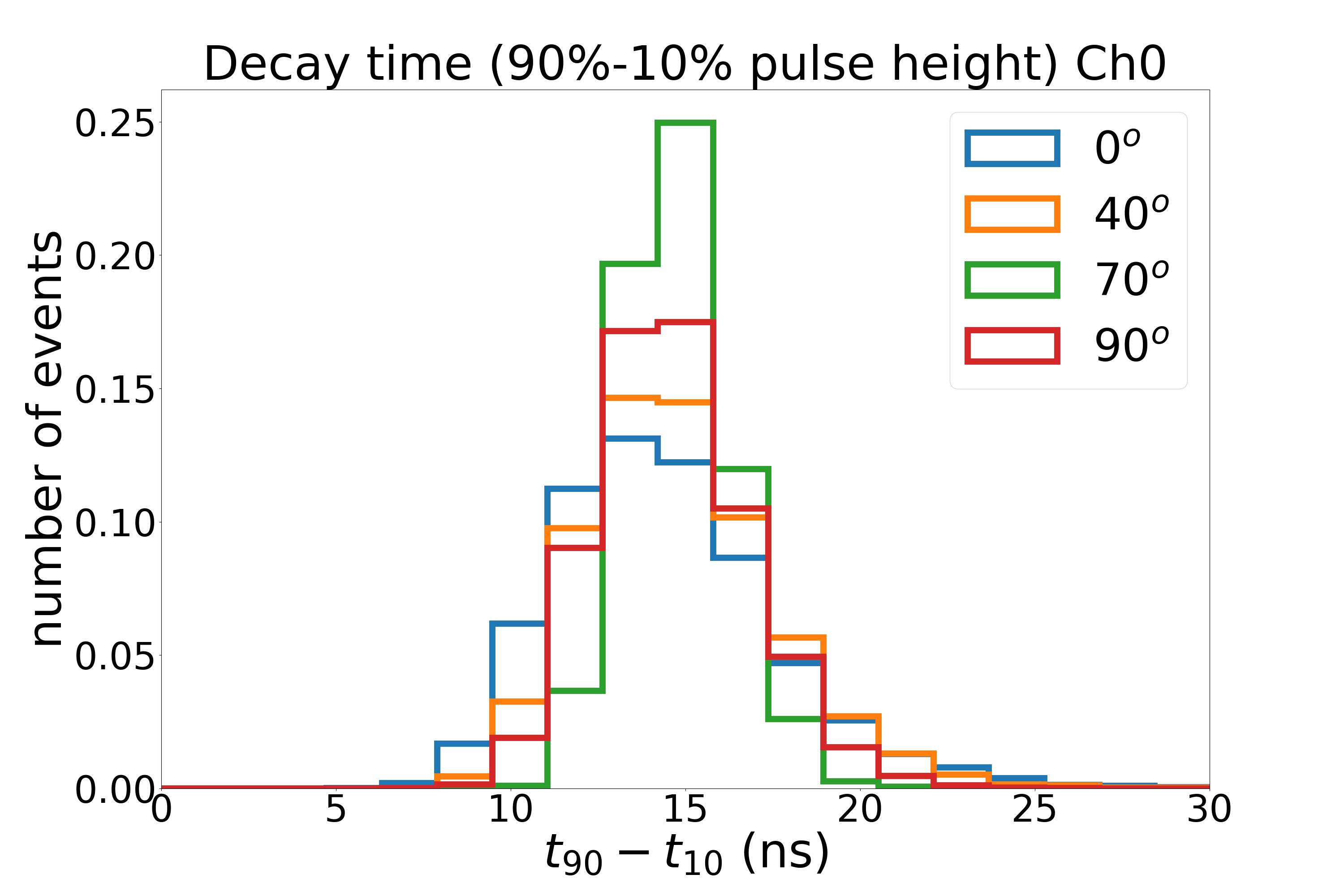} \label{fig:decay_time_ch0} } 
    \caption{Rising and decay times (time difference between 10\% and 90\% of the pulse height) for channel 0 at different rotation angles.}
    \label{fig:rising_decay_time}
\end{figure}

Comparison of the rising time and decay times between four different rotation angles are shown in Fig.~\ref{fig:rising_decay_time}. The decay time (approximately 15 ns) is found to be independent of the rotation angle, dominated by the \sipm\ and electronics responses. In contrast, the rising time (approximately 2 ns) remains relatively stable from 0 to $\pm 70^{\circ}$ but begins to differ beyond $\pm 70^{\circ}$.

\begin{figure}[bt]
    \centering
     \includegraphics[width=0.60\textwidth]{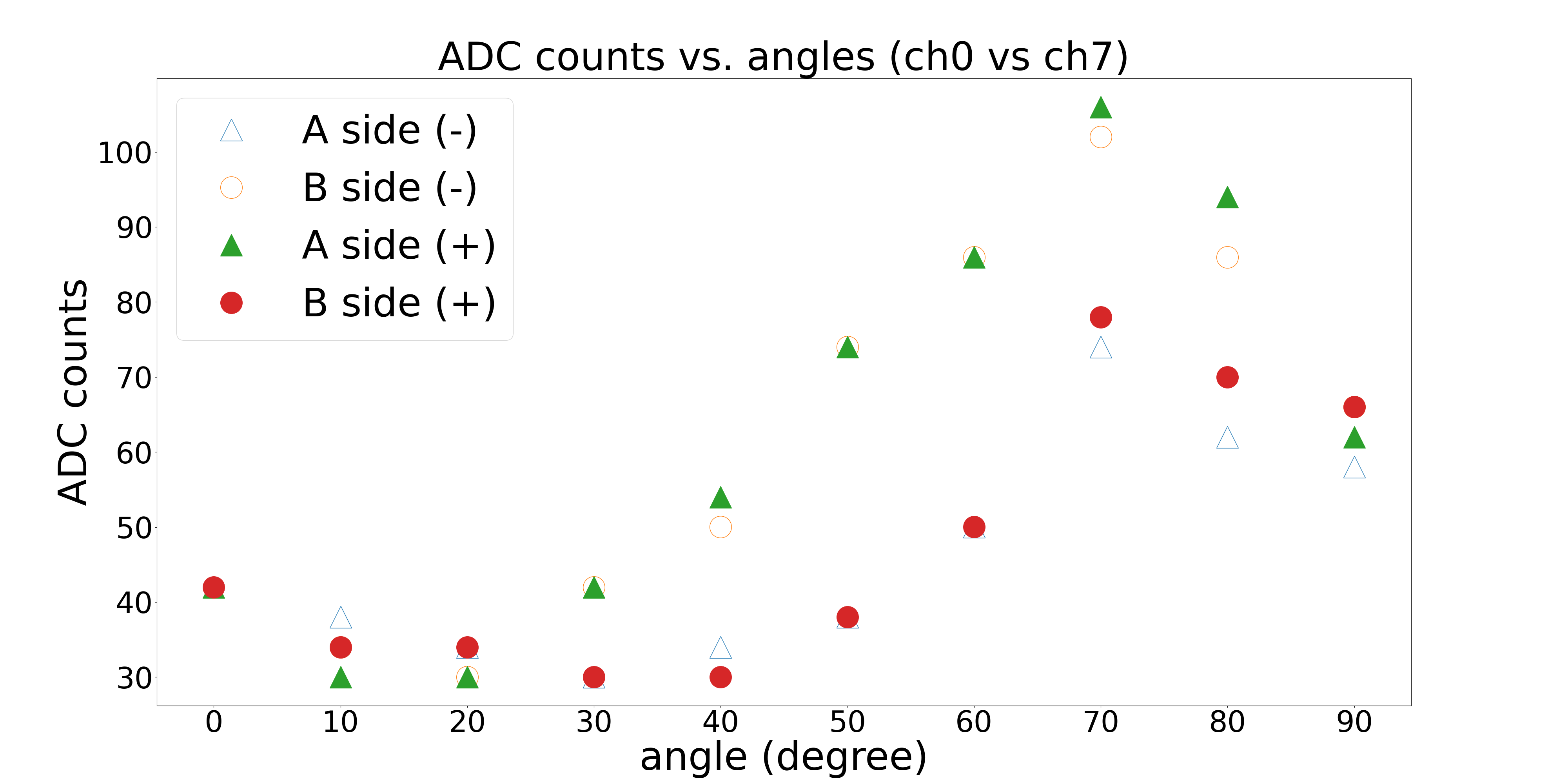}
    \caption{Average ADC value for channels 0 (A side) and 7 (B side). Data acquired with negative rotation angles are mirrored to positive rotation angles. Points with the label ``A side (+)" mean data taken with positive rotation angles ($+\theta$) for channel 0, and points with the label ``A side (-)" mean data taken with negative rotation angles ($-\theta$) for channel 0. For the $x$ axis, the absolute value of the rotation angle is shown.}
    \label{fig:ch_comparison_angle}
\end{figure}

Due to the geometric symmetry in our experimental setup, channels 0 and 7, 1 and 6, 2 and 5, and 3 and 4, are mirror images of each other. Consequently, we anticipate similar responses when the rotation angle is flipped. Figure~\ref{fig:ch_comparison_angle} illustrates the average ADC value as a function of the rotation angle for channels 0 and 7. Data acquired with negative rotation angles are mirrored to positive rotation angles. It is evident that the data obtained with $-\theta$ for channel 0 closely resemble the data obtained with $+\theta$ for channel 7. Similarly, the data obtained with $-\theta$ for channel 7 closely resemble the data obtained with $+\theta$ for channel 0.

\begin{figure*}
    \centering
     \subfloat[]{\includegraphics[width=0.33\textwidth]{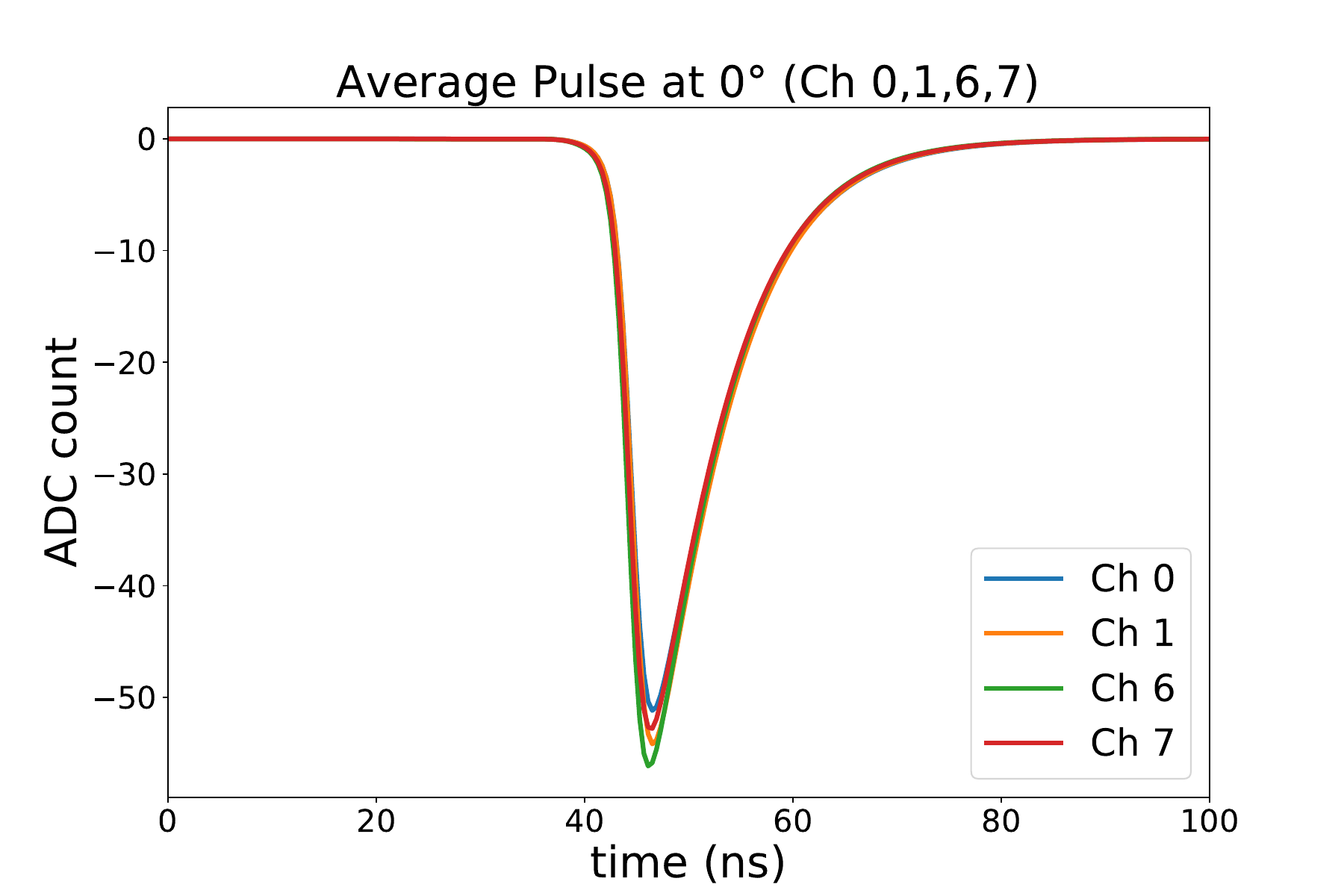} 
     \label{fig:average_timing_spectrum_0} } 
     \subfloat[]{\includegraphics[width=0.33\textwidth]{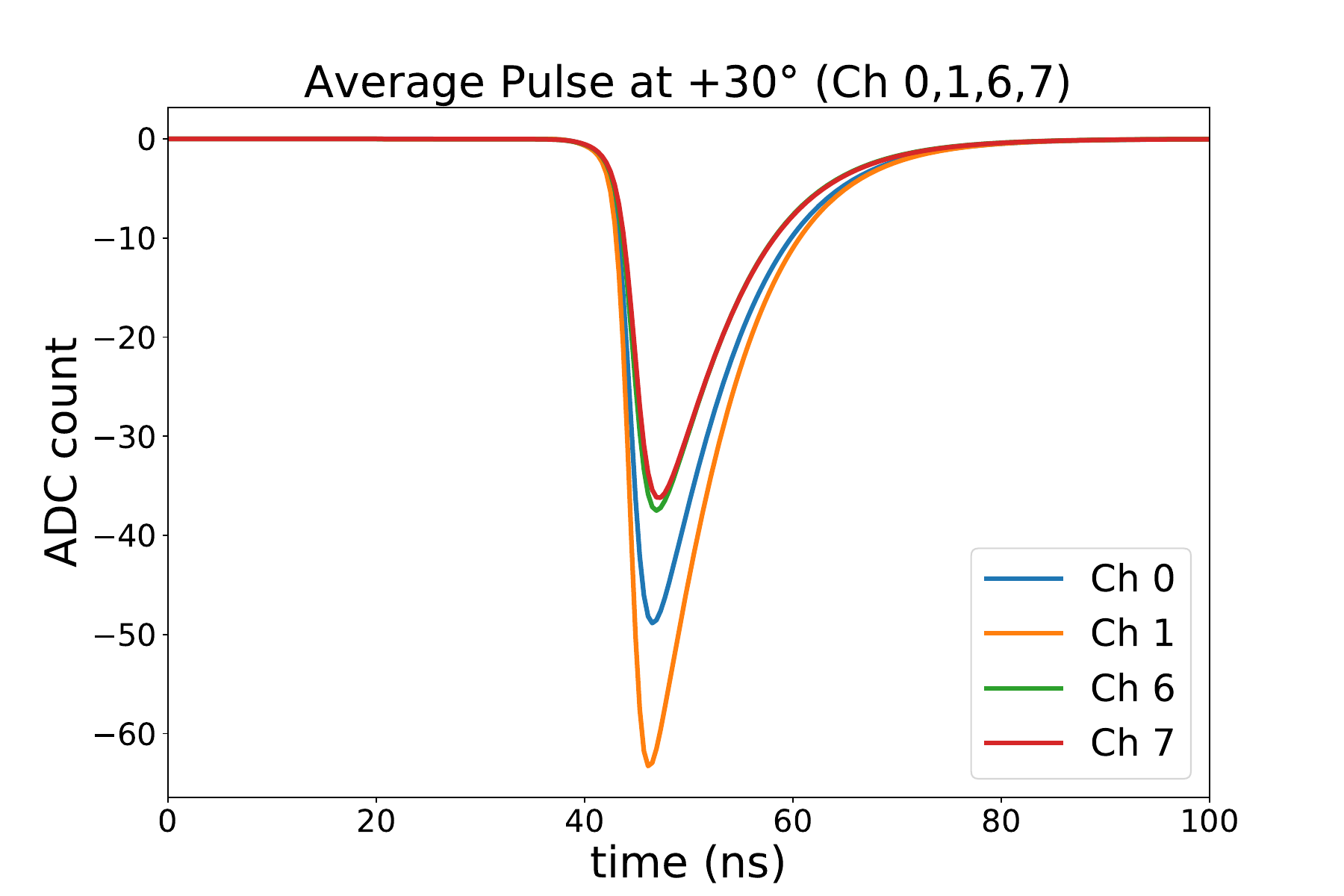} 
     \label{fig:average_timing_spectrum_p30} } 
     \subfloat[]{\includegraphics[width=0.33\textwidth]{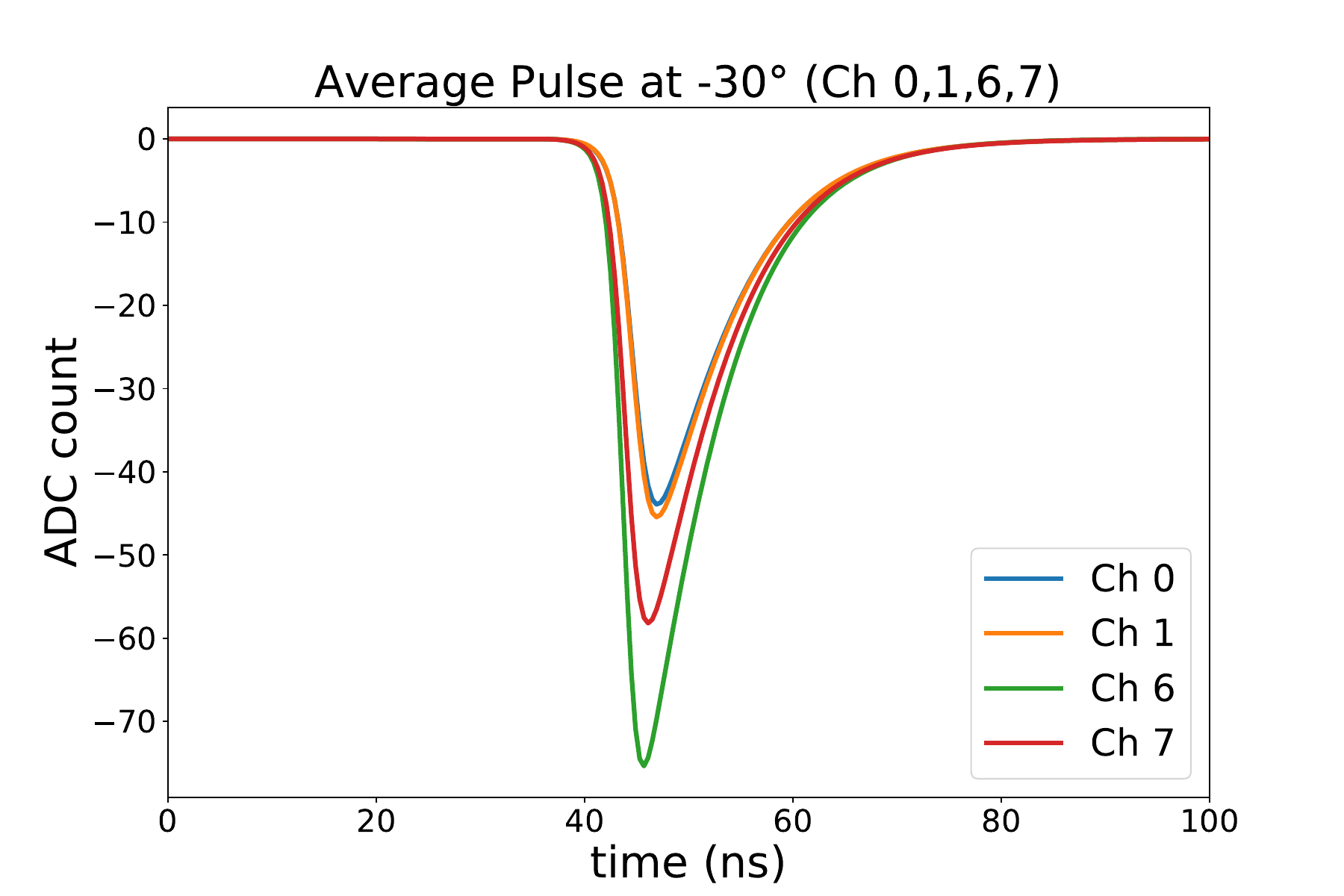} 
     \label{fig:average_timing_spectrum_n30} } 
    \caption{Average timing spectra at three different rotation angles: (a) 0$\degree$, (b) $+30\degree$, and (c) $-30\degree$.}
    \label{fig:average_timing_spectrum}
\end{figure*}

Figure~\ref{fig:average_timing_spectrum} illustrates the average timing spectra for four representative channels (0 and 1 on side A, and 6 and 7 on side B) at three different rotation angles: $0^{\circ}$, $-30^{\circ}$, and $+30^{\circ}$. The peak amplitudes are comparable across all four channels at $0^{\circ}$, indicating a similar count of Cherenkov photons reaching both ends of the crystal. In addition, on average channels 1 and 6 detect approximately two more Cherenkov photons than channels 0 and 7, consistent with expectations from the simulation. This difference primarily arises from the relative positions of these \sipms. For $\theta=+30^{\circ} (-30^{\circ})$, we anticipate more Cherenkov photons reaching side A (B) compared to side B (A), which aligns with the observed data in Fig.~\ref{fig:average_timing_spectrum}.

\begin{figure}[bt]
    \centering
    \includegraphics[width=0.4\textwidth]{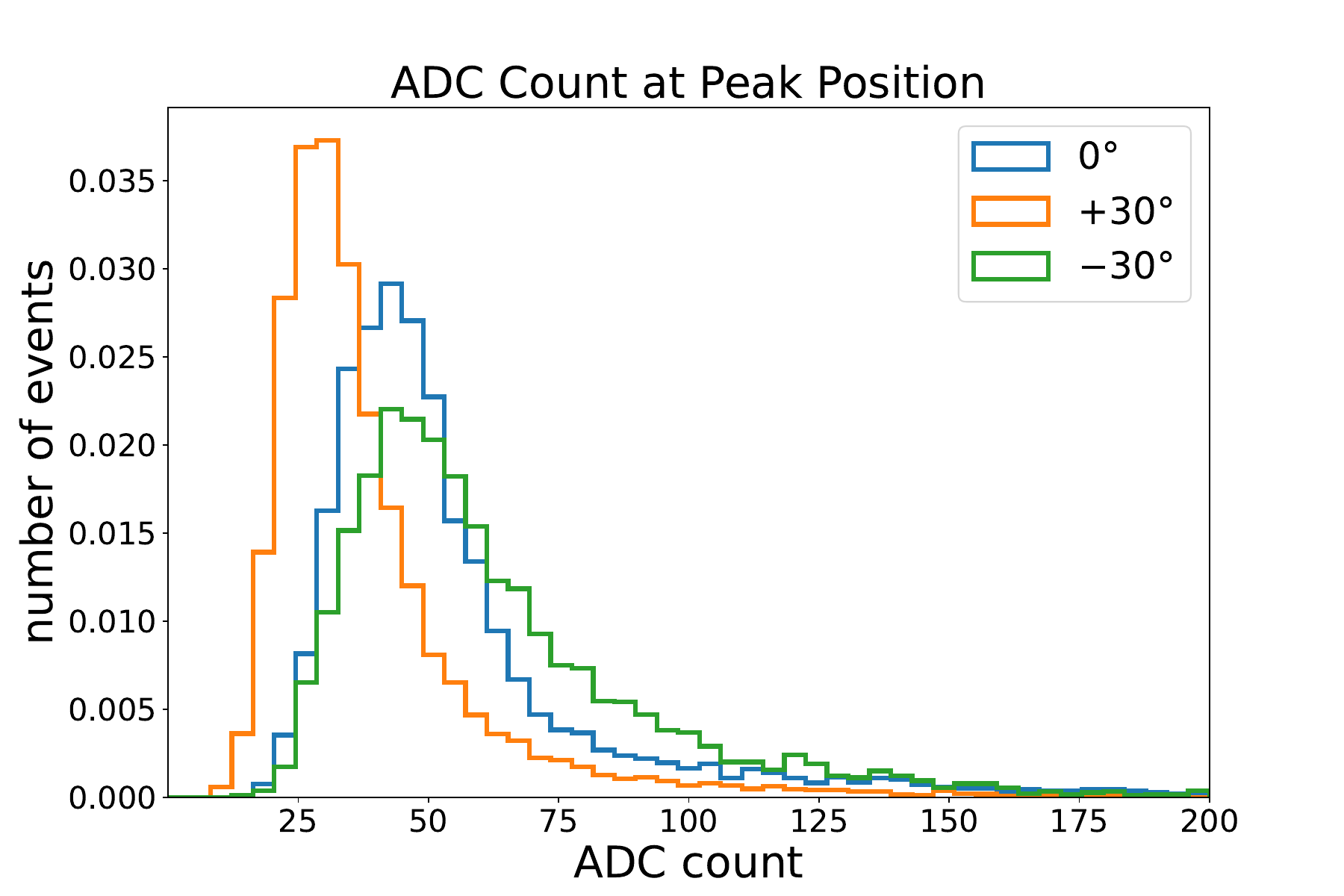} \label{fig:fitted_amplitude_ch7} 
    \caption{Fitted ADC count at the peak position for $\theta=0^{\circ}$, $-30^{\circ}$, $+30^{\circ}$ (channel 7).}
    \label{fig:fitted_amplitude_vs_angle}
\end{figure}

The ADC count at the peak position reflects the number of detected Cherenkov photons. For each event, the timing spectrum for each \sipm\ is fitted to the function described by Eqn.~\ref{eqn:fit_eqn}. The distributions of the fitted ADC count at the peak position for channel 7 at three rotation angles are displayed in Fig.~\ref{fig:fitted_amplitude_vs_angle}. The average number of Cherenkov photons detected by this \sipm\ in channel 7 is approximately 10 at $+30^{\circ}$, increasing to about 15 at $0^{\circ}$ and $-30^{\circ}$.

\begin{figure*}[ht]
    \centering
    \subfloat[]{\includegraphics[width=0.35\textwidth]{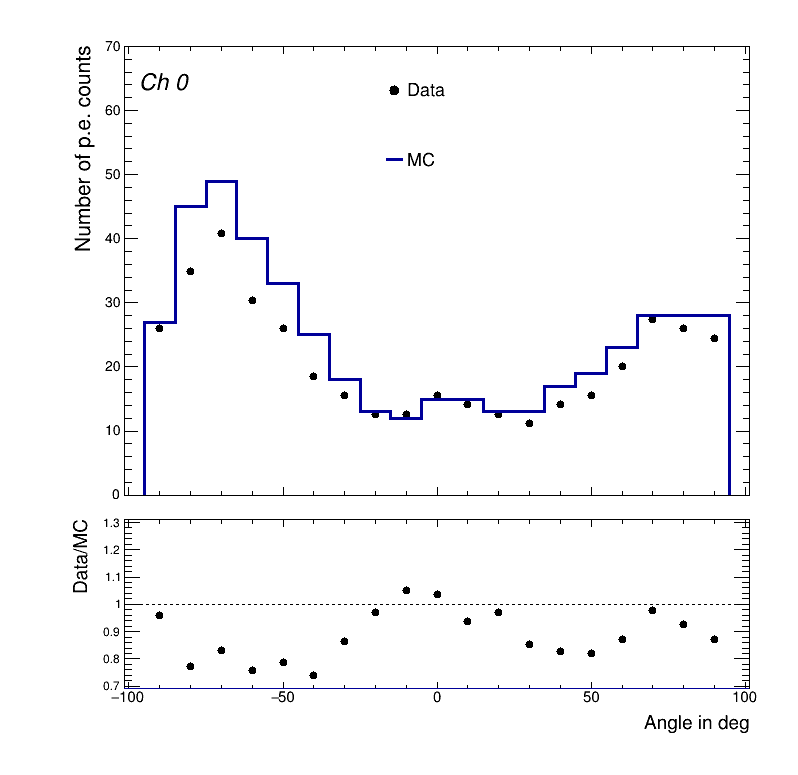} \label{fig:comparison_ch0} } 
    \subfloat[]{\includegraphics[width=0.35\textwidth]{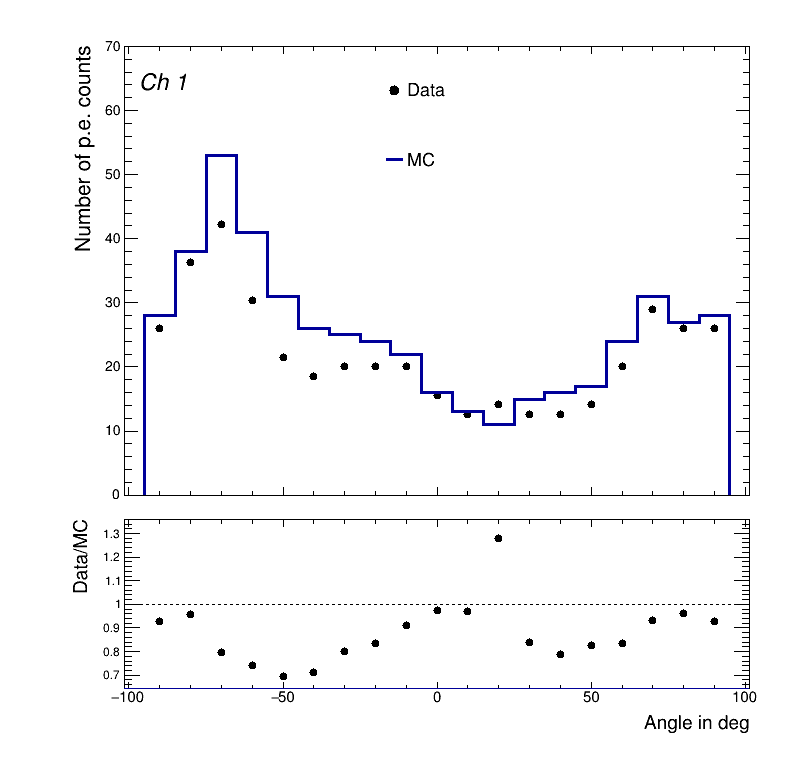} \label{fig:comparison_ch1} } \\
    \subfloat[]{\includegraphics[width=0.35\textwidth]{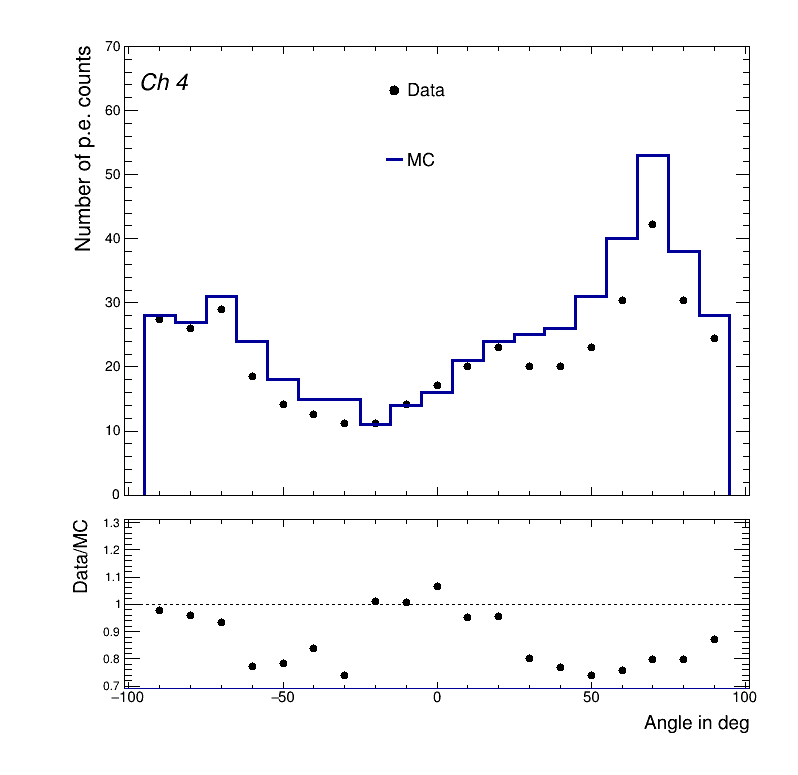} \label{fig:comparison_ch4} } 
    \subfloat[]{\includegraphics[width=0.35\textwidth]{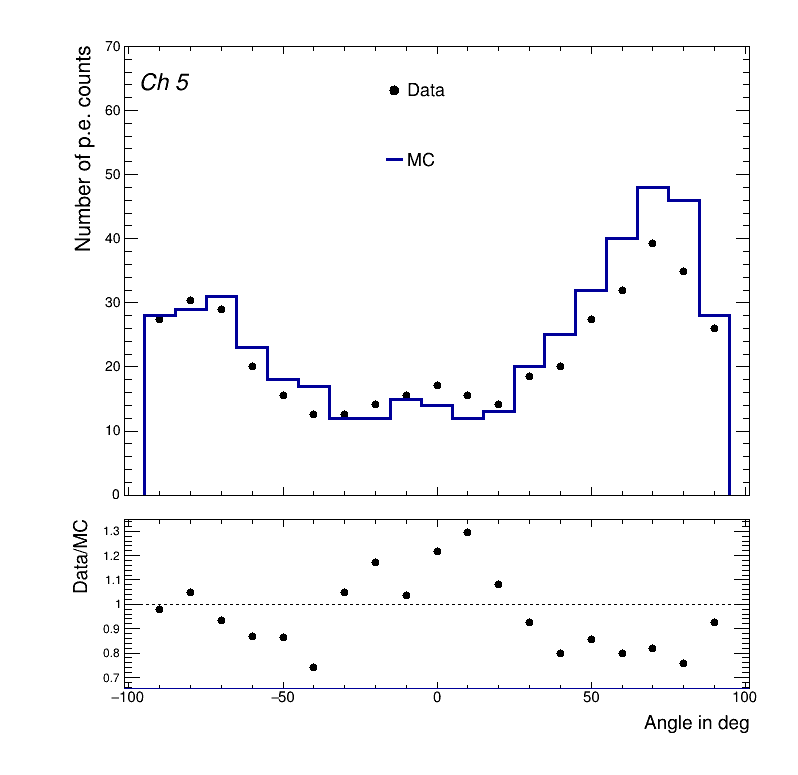} \label{fig:comparison_ch5} } 
     \caption{Comparison between data and simulation for the mean number of photons detected for channel 0 (a), channel 1 (b), channel 4 (c), and channel 5 (d). Each bottom panel shows the ratio of the mean number of photons measured in data and in the simulation.}
    \label{fig:pbf}
\end{figure*}

Figure~\ref{fig:pbf} shows the most probable value of the fitted peak amplitude distribution (indicating the average number of detected photons) as a function of the rotation angle $\theta$ for four channels (0, 1, 4, and 5). The data are compared to the simulation results, where the average ADC values measured in the data are converted into the numbers of photons detected. The simulation reproduces the general trends observed in the data as a function of the rotation angles for different SiPMs. However, the absolute number of photons detected can differ between the data and simulation by 30\% for certain rotation angles. In general, the number of detected photons increases with increasing $|\theta|$. However, this trend reverses at $\theta \simeq 70^{\circ}$, corresponding to the proton's path length reaching its maximum at $\theta=67.4^{\circ}$. 

As mentioned in Section~\ref{sec:simulation}, the settings for the optical properties of the wrappers and interfaces significantly impact the degree of agreement between data and the simulation results. A significant difference in counts is observed for the near-side channels in the simulation when the surface is set to 0\% lobe compared to when it is set to a non-zero lobe percentage, especially for $|\theta|<20^{\circ}$. In general, as the value of the lobe percentage increases, the angular dependence curve becomes smoother for $|\theta|<70^{\circ}$ and the count difference between the correlated pairs of channels for $|\theta| \sim 20^{\circ}$ drops from more than 15 photons for the 0\% lobe case to about 5 photons for the 100\% lobe case. Similarly, we also ran simulations over several values of the roughness parameter sigma-alpha, keeping the proportions of lobe and spike reflection types at 25\% and 75\%. For a rough surface, the angle of incidence and reflection for a particular track is calculated by considering that the normal at the point of incidence is not necessarily the same as the average surface normal. We define alpha as the angle between the normal of the micro facet (which is basically each point of the rough surface) and the average surface normal, and sigma as the standard deviation for the distribution of alpha (which is assumed to be Gaussian). This means that the more the value of sigma-alpha, the rougher the surface. Also a large difference is observed between the perfectly smooth case, and a case where the sigma-alpha parameter is non-zero, but there is not much difference in the counts among the runs with non-zero sigma-alpha values.

\begin{figure*}[ht]
    \centering
    \subfloat[]{\includegraphics[width=0.4\textwidth]{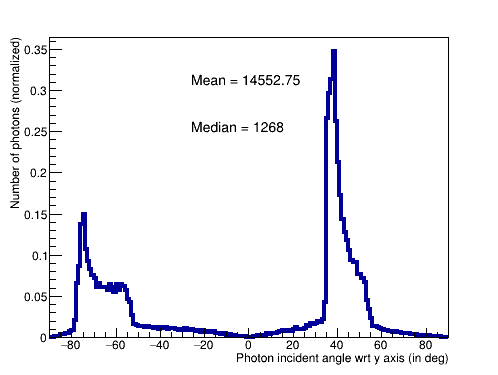} \label{fig:PhotonAngles_20degree_Cut1} } 
    \subfloat[]{\includegraphics[width=0.4\textwidth]{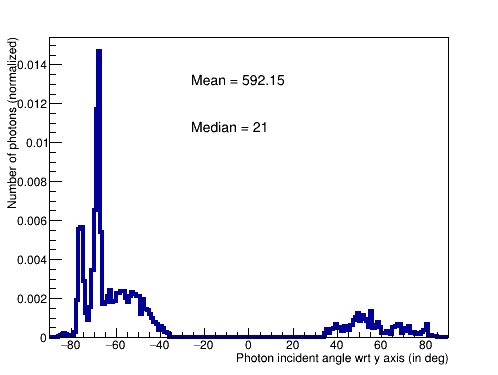} \label{fig:PhotonAngles_20degree_Cut3} } 
    \caption{(a) Incident angles of photons hitting the top surface at their first reflection; and (b) Incident angles of photons that reach the SiPM after passing through the silicone cookie. The crystal rotation angle is $+20^{\circ}$.}
    \label{fig:PhotonAngles_20degree}
\end{figure*}

An intriguing observation is the reduction in the number of detected photons when the rotation angle is near $20^{\circ}$. Detailed simulations are conducted for $\theta=20^{\circ}$ and $\theta=10^{\circ}$ to investigate this phenomenon. 
For the case with $\theta=20^{\circ}$, according to geometrical optics, the incident angle of photons striking the long surface is concentrated around $36^{\circ}$ and $76^{\circ}$, as confirmed by the simulation in Fig.~\ref{fig:PhotonAngles_20degree}(a). Subsequently, in Fig.~\ref{fig:PhotonAngles_20degree}(b), the same angles are shown for photons reaching side B of the crystal and detected by the corresponding \sipms. These photons have an incident angle of around $14^{\circ}$ or $54^{\circ}$ on the exiting surfaces. Given that the total internal reflection angle from the crystal to the silicone rubber is $51.8^{\circ}$, many of the photons with an incident angle close to $54^{\circ}$ will be reflected and thus not collected by the \sipms. This explains the observed decrease in Cherenkov photons when $\theta$ is close to $20^{\circ}$.

\begin{figure*}[ht]
    \centering
    \subfloat[]{\includegraphics[width=0.4\textwidth]{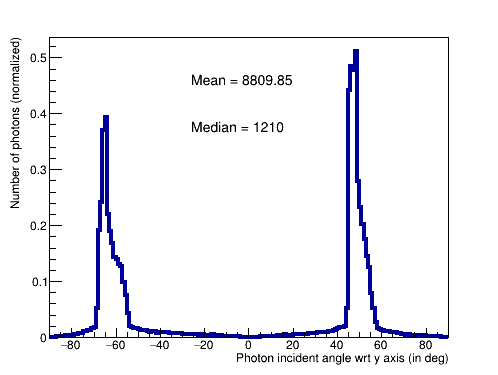} \label{fig:PhotonAngles_10degree_Cut1} } 
    \subfloat[]{\includegraphics[width=0.4\textwidth]{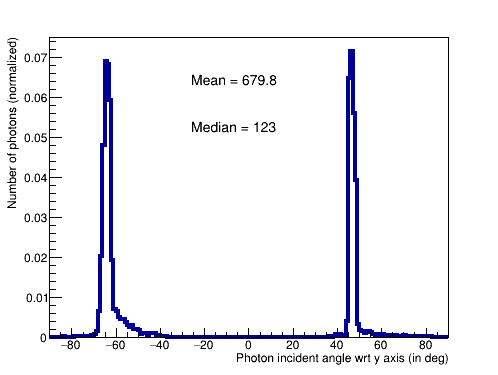} \label{fig:PhotonAngles_10degree_Cut3} } 
    \caption{(a) Incident angles of photons hitting the top surface at their first reflection; and (b) Incident angles of photons that reach the SiPM after passing through the silicone cookie. The crystal rotation angle is $+10^{\circ}$.}
    \label{fig:PhotonAngles_10degree}
\end{figure*}

For the case with $\theta=10^{\circ}$, Fig.~\ref{fig:PhotonAngles_10degree} depicts the corresponding incident angles on the long surface for these photons with their first reflections and detection by the \sipms. The majority of these photons have an incident angle of around $46^{\circ}$ or $66^{\circ}$ on the long surface and an incident angle of around $24^{\circ}$ or $44^{\circ}$ on the exiting surfaces. These photons do not undergo internal reflection between the surfaces of the crystal and the silicone rubber and are detected by the \sipms.

\begin{figure}[bt]
    \centering
     \includegraphics[width=0.5\textwidth]{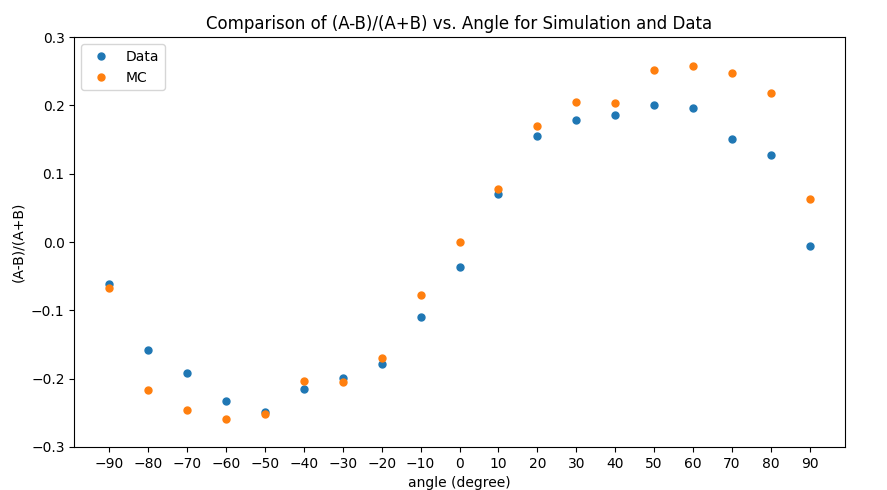}
    \caption{Comparison of A-B asymmetry for simulation and data at different rotation angles.}
    \label{fig:asymmetry}
\end{figure}

The asymmetry in the integrated ADC counts between the A and B sides (referred to as A-B asymmetry) is calculated for each rotation angle, as depicted in Fig.~\ref{fig:asymmetry}. The asymmetry is defined as $(A-B)/(A+B)$, where $A$ and $B$ represent the sum of integrated ADC counts for all channels from each side, respectively (excluding channel 2). An angle dependence in the asymmetry is observed, showing an opposite trend for positive and negative rotation angles. This behavior is expected due to the directionality of the Cherenkov photons. The maximum asymmetry is approximately 20\%, occuring when the rotation angle is close to $\pm 50\degree$. The asymmetry measured in data is also compared to the predictions from the simulation, showing reasonable agreement. 

\begin{figure*}[ht]
    \centering
     \includegraphics[width=0.80\textwidth]{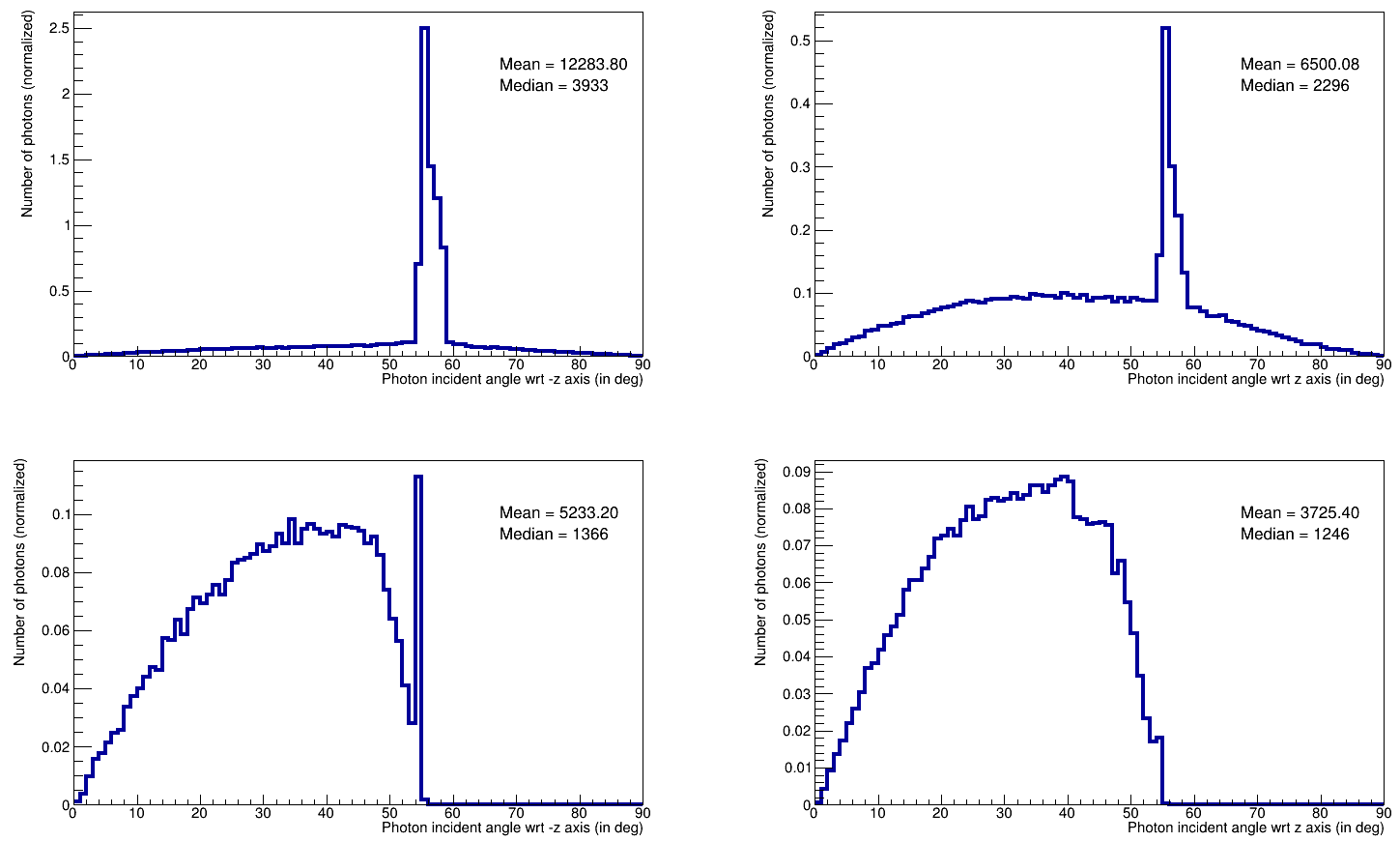}
    \caption{Top left: incident angles of photons hitting side A; bottom left: incident angles of photons detected by the \sipms\ on side A; top right: incident angles of photons hitting side B; bottom right: incident angles of photons detected by the \sipms\ on side B. The crystal rotation angle is $+90^{\circ}$.}
    \label{fig:angle_90degree}
\end{figure*}

It is also interesting that the asymmetry is close to 0 when $\theta=\pm 90^{\circ}$, indicating that the numbers of Cherenkov photons detected at the opposite ends of the crystal are similar. Detailed simulation studies are conducted for the case with $+90^{\circ}$. Figure~\ref{fig:angle_90degree} illustrates the photon incident angle distributions for all photons reaching side A (top left plot) and those detected by the \sipms\ on side A (bottom left plot). The figure also shows the corresponding distributions for all photons reaching side B (top right plot) and those detected by the \sipms\ on side B (bottom right plot). It can be observed that most generated photons reaching the two sides have incident angles concentrated around $56^{\circ}$, which is close to the internal reflective angle of $51.8^{\circ}$. Consequently, most photons will be reflected and not detected by the \sipms. The number of photoelectrons detected by the \sipms\ on side A is similar to the number of photoelectrons detected by the \sipms\ on side B; as a result, the asymmetry is close to 0 when $\theta=\pm 90 ^{\circ}$.

\section{Conclusions}
A study of the \cherenkov~light yield for a \pbf~crystal excited by protons as a function of the proton's incident angle is presented. A non-trivial angular dependence is observed in data for all SiPMs, and the general trend can be understood by considering the crystal's geometry, contributions of index matching, total internal reflection, and non-specular reflections of the crystal, the silicon rubber, and the wrapper material. The simulation can reproduce data for the number of photons detected within 30\%. Attention to these aspects will be crucial for realizing the CalVision program~\cite{calvision}. The setup we have will be useful for future DR studies using \pbwo\ and \bgo\ crystals.

\section{Acknowledgments}
The authors would like to thank the staff at Fermilab's test beam facility. 
 This work was supported in part by U.S. Department of Energy Grant DE-SC0022045.

\bibliography{main}

\end{document}